\documentclass[aip,jcp,amsmath,amssymb,preprint]{revtex4-2}
\usepackage{graphicx}% Include figure files
\usepackage{dcolumn}% Align table columns on decimal point
\usepackage{bm}% bold math
\usepackage{hyperref} %hyper reference
\hypersetup{colorlinks=true,
	linkcolor=blue,
	anchorcolor=blue,
	citecolor=blue}

\begin{document}

%\preprint{AIP/123-QED}

\title{The Real Space Correlation Function of Gaussian Chain in Spin-echo Small Angle Neutron Scattering} %Title of paper



\author{Tengfei Cui}
\email[]{cuitengfei20@gscaep.ac.cn}
%\thanks{}
\affiliation{Department of Nuclear Science and Technology, Graduate School of China Academy of Engineering Physics, Beijing 100193, China}

\author{Xiangqiang Chu}
\email[]{xiangchu@cityu.edu.hk}
%\thanks{}
\affiliation{Department of Physics, City University of Hong Kong, Hong Kong 999077, China}


\date{\today}

\begin{abstract}
The utilization of spin-echo small angle neutron scattering (SESANS) for the analysis of structures in soft matter is becoming increasingly prevalent. In this context, the Gaussian chain model and the corresponding framework for calculating the theoretical SESANS correlation function are presented briefly. This work provides a novel theoretical derivation to analytically obtain the density-density correlation function and the SESANS spatial correlation function of Gaussian chains. Furthermore, we discuss the relationship between these correlation functions and the measured neutron polarizability of SESANS. These functions possess favorable mathematical properties, rendering them directly suitable for the analysis and fitting of data obtained from polymer samples via SESANS.
\end{abstract}


\maketitle %\maketitle must follow title, authors, abstract

\section{INTRODUCTION}%\label{}
Flexible and amorphous string-like polymer chains which are in a dilute solution solvent or molten are similar to random coils. It is appropriate for these chains to be well characterized by the ideal chain model. For the ideal chains, there is no long-range interaction between any two monomers and their conformations are generated by random walks of chains, whose end-to-end vectors conform a Gaussian distribution.\cite{kawakatsu2004statistical} Fortunately, the Gaussian chain model is a good mathematical representation of the properties of an ideal chain because the overall statistical properties do not depend on the local details of a chain. As is known, we can use small angle X-ray scattering (SAXS) or small angle neutron scattering (SANS) experiments getting the form factor $ F(Q) $ to characterize the physical properties of polymer molecules. $ Q $ is the wave vector transfer and it means the length of a reciprocal space. And the form factor is an important function for describing the shape of a polymer chain. In contrast, there is a real-space technique called Spin-echo small-angle neutron scattering (SESANS) which measures the projection of the density–density correlation function of a sample, rather than, as in SAXS and SANS, its Fourier transform. Thus it presents an alternative method to SANS for probing soft matter structure including polymers.

SESANS was developed for the direct structural determination at large length scales ranging from the nano- to the micro-meter domain\cite{bouwman2000development}. This technique is based on the Larmor precession of polarized neutrons in magnetic fields configured to encode the neutron momentum transfer and accordingly it measures neutrons polarization $ P(z) $ that gives the spatial correlation of probed materials as a function of the spin-echo length $ z $. The spatial correlation is denoted as $ G(z) $ obtained from polarization $ P(z) $. It is the SESANS real-space correlation function. Besides, it is important to note that in comparison with the small angle coherent scattering cross section, which is known to be the Fourier transform of the two-point spatial density-density correlation $ \gamma(r) $, namely, the Debye correlation function, $ G(z) $ is found to relate with $ \gamma(r) $ via an Abel transformation due to the action of spin echo.

In the last two decades, SESANS technology has been fully developed and applied to several structural investigations of soft matter systems, especially colloids. In this paper, we derive the Debye correlation function $ \gamma(r) $ from the significant properties of the Gaussian chain model. Then we obtain the spatial correlation function $ G(z) $ by Abel transform and the polarization function $ P(z) $ of SESANS.

\section{GENERAL DISCUSSION OF THE GAUSSIAN CHAIN AND SESANS}
Gaussian chain is a widely used polymer chain model. In subsection \ref{Gaussianchain}, we will briefly introduce the Gaussian chain model and its application to the SAXS and SANS. In subsection \ref{SpatialcorrelationfunctionofSESANS}, we will explore what information we can get from experiments by the SESANS technique and how we relate these data to physical models.

\subsection{Gaussian Chain}\label{Gaussianchain}
\begin{figure}
	\centering
	\includegraphics[width=0.4\linewidth]{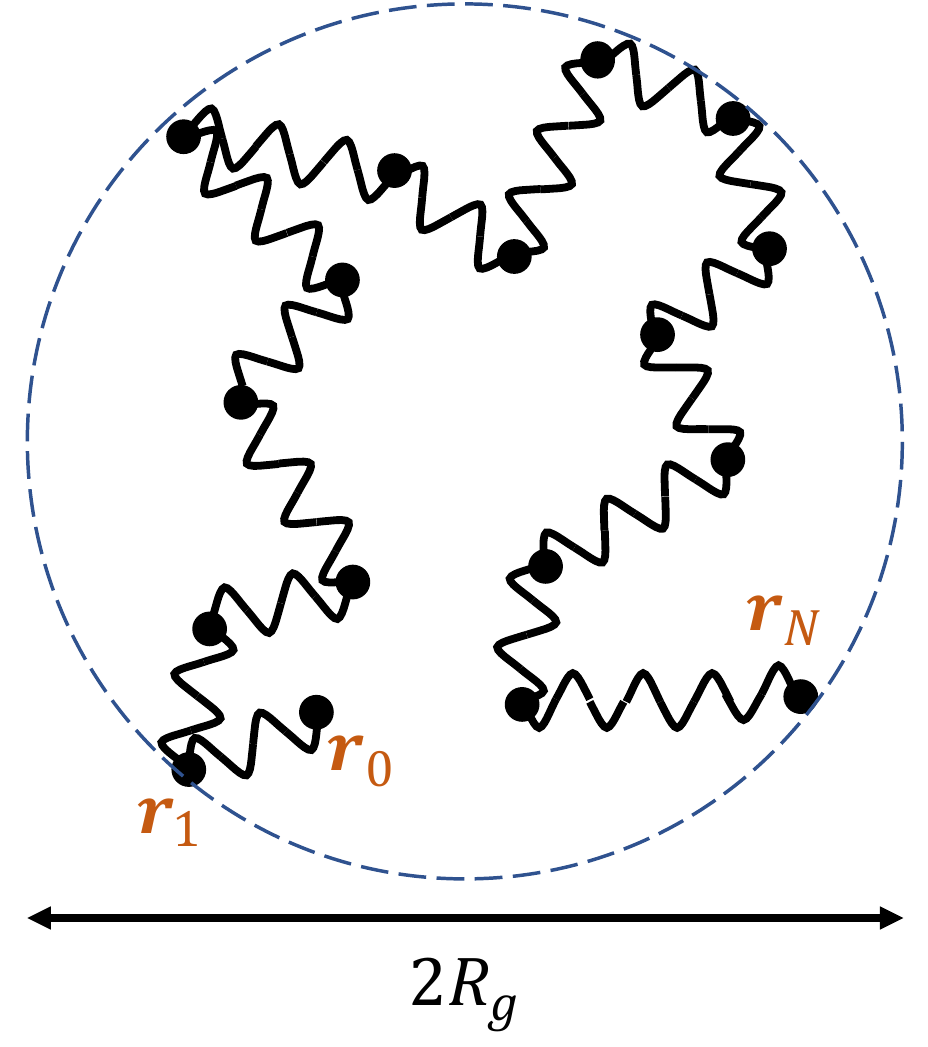}
	\caption[]{Gaussian chain.}
	\label{fig:fig1}
\end{figure}

Gaussian chains are usually represented by $ N+1 $ beads connected by harmonic spring (Fig.\ref{fig:fig1}). Beads can be considered as points, because we assume that the size of the unit of the molecular model is much larger than the atomic scale. And a Gaussian chain whose bond length has the Gaussian distribution\cite{doi1988theory}
\begin{eqnarray}\label{Gaussiandistribution}
	\psi(\bm{r}) = \left( \frac{3}{2\pi b^2} \right) ^{\frac{3}{2}} \exp \left( -\frac{3\bm{r}^2}{2b^2} \right),
\end{eqnarray}
so that $ \left\langle \bm{r}^2 \right\rangle = b^2 $, where $ b $ is the effective bond length. Besides, the distribution of the bond length $\left|  \bm{r}_n - \bm{r}_m \right|  $ between any two units $ n $ and $ m $ in this chain is Gaussian
\begin{eqnarray}\label{twounitsdistribution}
	\Phi(\bm{r}_n-\bm{r}_m) = \left(\frac{3}{2\pi b^2 \left| n-m \right| } \right)^{\frac{3}{2}}  \exp \left[   -\frac{3(\bm{r}_n-\bm{r}_m)^2}{2\left| n-m \right| b^2} \right] .
\end{eqnarray}

Moreover, in the SAXS and SANS experiment, the structure factor of the Gaussian chain $ S(Q) $ is proportional to form factor $ F(Q) $, which means  that the system of Gaussian chains are fractal. And we know that
\begin{eqnarray}\label{Debyefunction}
	F(Q) = \frac{2 \left[  \exp(-Q^2R_g^2)+Q^2R_g^2 -1 \right]  }{Q^4R_g^4}
\end{eqnarray}
is named Debye function, and $ R_g $ is the radius of gyration. It is worth noting that $ F(Q) $ is only a function of the radial direction of $ \bm{Q} $. That is to say a sample consisting of large numbers of Gaussian chains is isotropic.

We can obtain $ \gamma(r) $ by the Fourier transform of $ F(Q) $ in principle. Consequently this $ \gamma(r) $ at distances $ r $ smaller than the chain size $ R_{g} $ decreases like $ 1/r $ which pointed out by De Gennes.\cite{de1979scaling} However, $ \gamma(r) $ is divergent as $ r $ tends to 0. This leads to an infinite correlation length $ \xi $ within the Abel transform of $ \gamma(r) $. So this function $ \gamma(r) $ cannot be used directly in the data analysis of SESANS. The fundamental reason for this result is that the form factor $ F(Q) $ in Eq.(\ref{Debyefunction}) is an approximate analytic expression. Therefore, in the next section we will introduce a novel derivation to obtain the Debye correlation function $ \gamma(r) $ of the Gaussian chain model.

\subsection{Spatial Correlation Function of SESANS}\label{SpatialcorrelationfunctionofSESANS}
The  working mechanism principle of SESANS has been widely reported. Here we briefly introduce the following fundamental principles based on the definition of Kruglov. The SESANS instrument can measure the variation of the neutron polarizability $ P(z) $, where $ z $ is the  spin-echo length. It can be analytically expressed as a one-dimensional real space correlation function $ G(z) $ with approximations. Let us consider a sample with thickness $ t $, illuminated by a neutron beam of cross section $ S $, and the fixed wavelength of neutrons is $ \lambda $, so $ V = tS $ is the volume of the sample illuminated. The relationship between $ P(z) $ and $ G(z) $ is written as\cite{krouglov2003real}
\begin{eqnarray}\label{P}
	P(z) = \exp\left\lbrace \frac{\lambda ^2}{S} \left[  G(z)-G(0) \right]  \right\rbrace  = \exp \left\lbrace \frac{\lambda ^2}{S} G(0) \left[  G_0(z)-1 \right]  \right\rbrace ,
\end{eqnarray}
where, $ G_0(z) = G(z)/G(0) $. We can easily get $ P(z) $ equals to 1 at $ z = 0 $ from this equation.

The correlation function $ G(z) $ is related to a Debye-type correlation function $ \gamma(\bm{r}) $, defined as
\begin{eqnarray}
	\gamma(\bm{r}) =  \int_V \Delta \rho(\bm{r}') \Delta \rho (\bm{r}'+\bm{r} ) ~ d\bm{r}',
\end{eqnarray}
here $ \Delta \rho (\bm{r}')$ is the difference between the neutron scattering length density (SLD) averaged over the whole sample and the the local SLD at position $ \bm{r}' $.\cite{kruglov2005correlation} For an isotropic density distribution, Rekveldt first pointed out that $ G(z) $ can be related with $ \gamma(r) $ by the Abel transformation
 \begin{eqnarray}\label{G}
	G(z) = 2\int_{z}^{\infty} \frac{\gamma(r)r}{\sqrt{r^2-z^2}} dr 
\end{eqnarray}
and $ \xi = G(0) = 2\int_{0}^{\infty} \gamma(r) ~ dr $. Here $ \xi $ is called correlation length.\cite{andersson2008analysis}

\section{RESULTS AND DISCUSSION}
This section will present the correlation function of the density distribution $ \gamma(\bm{r}) $ for Gaussian coils, as well as the SESANS spatial correlation function $ G(z) $, and the SESANS neutron polarization rate $ P(z) $.

As is reported by Krouglov, for homogeneous particles, $ \gamma(r) $ has a clear physical significance. For example, the $ \gamma(r) $ of a spherical particle means the shared volume of this sphere and a phantom sphere shifted with respect to each other by the distance $ r $. But it can't be uesed for Gaussian coils because we have ignored the volume of monomers on each chain. What's more, the forms of chains are usually unfixed, so we have to average over all conformations of them. Therefore, $ \gamma(\bm{r}) $ takes the following form: \cite{li2010theoretical, krouglov2003structural} 
\begin{eqnarray}\label{3_eq_difine_gamma}
	\gamma(\bm{r}) = \left\langle \int_V \Delta \rho(\bm{r}') \Delta \rho (\bm{r}'+\bm{r} ) ~ d\bm{r}' \right\rangle
\end{eqnarray}	
$ \left\langle ~ \right\rangle  $denotes all conformations average.

\begin{figure}
	\centering
	\includegraphics[width=0.5\linewidth]{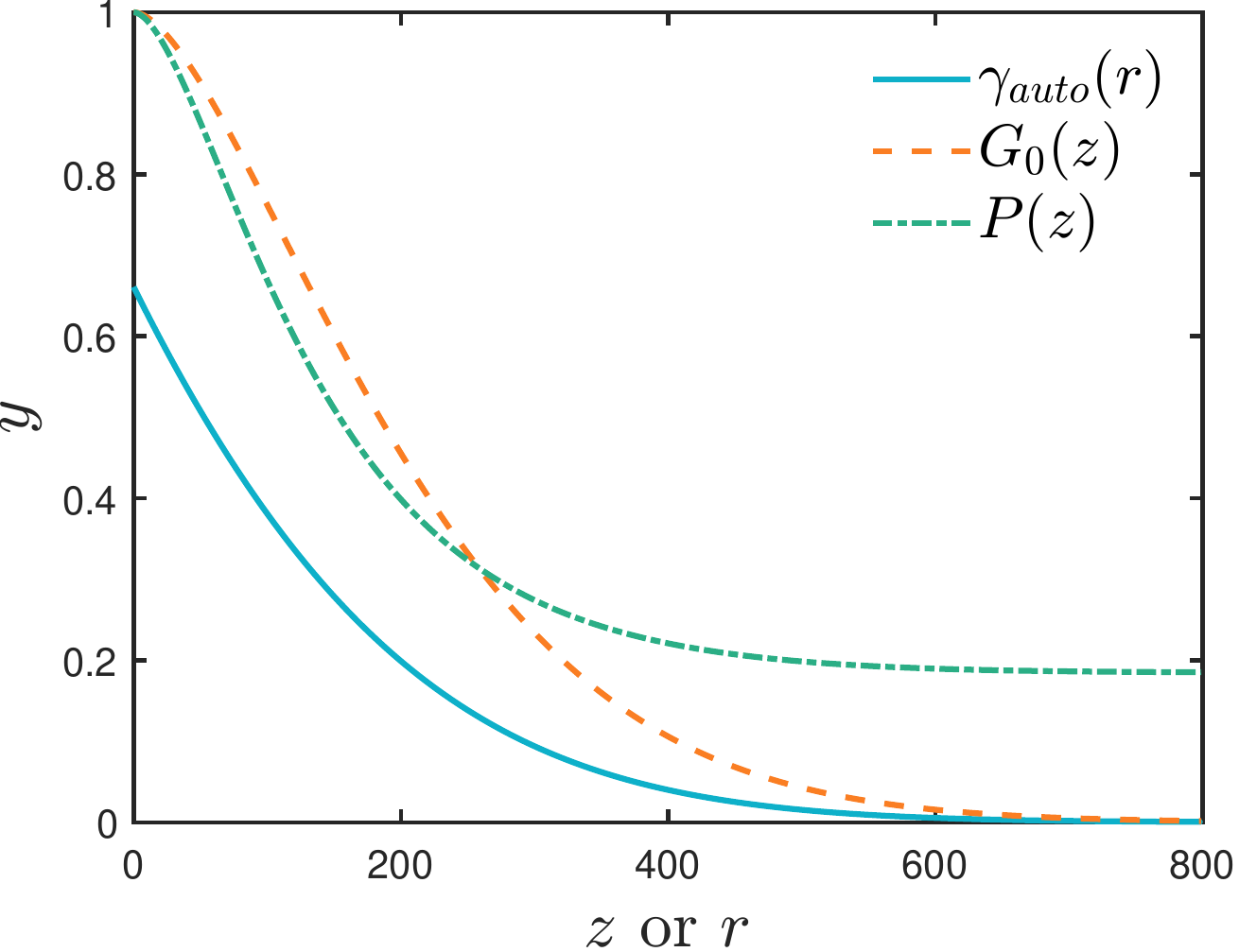}
	\caption[]{SESANS spatial correlation functions and neutron polarizability of the Gaussian chain with the radius of gyration $ R_g = 150 $. The solid line is the autocorrelation function $ \gamma_{auto}(r) $, the dashed line is the normalized SESANS spatial correlation function $ G_0(z) $, the dashed-dotted line is the neutron polarizability $ P(z) $.}
	\label{fig:fig2}
\end{figure}

% \ref{SpatialcorrelationfunctionofSESANS}
Now we consider a sample containing $ N_c $ polymer chains, where the percentage of deuterated chains is $ \phi $. And the scattering lengths of the hydrogenated monomer and deuterated monomer are $ b_H $ and $ b_D $ respectively.
We find that the $ \gamma(r) $ of Gaussian chains is proportional to the autocorrelation function of a single chain $ \gamma_{auto}(r) $ which is similar with form factor $ F(Q) $.
\begin{eqnarray}
	\gamma(r) =   A \gamma _{auto} (r)
\end{eqnarray}	
where, $ A =  N_c \phi (1-\phi)(b_H - b_D)^2 N^2/(4\pi R^3_g) $. Hence you could regard the $ \gamma(r) $ and $ \gamma_{auto}(r) $ in the real space as the $ S(Q) $ and $ F(Q) $ in the reciprocal space respectively. The detail of function $ \gamma_{auto}(r) $ (Fig.\ref{fig:fig2}) follows:
\begin{eqnarray}\label{gammaauto}
	\gamma _{auto}(r) = g(\eta) - \frac{1}{\sqrt{2}}~g\left( \frac{\eta}{\sqrt{2}} \right) - \eta ,
\end{eqnarray}
where $ \eta = r/(2R_g) $, and
\begin{eqnarray}\label{g}
	g(x) = \left( \frac{1}{x} + 2x  \right) {\rm erf}(x) + \frac{2}{\sqrt{\pi}} e^{-x^2} .
\end{eqnarray}	
$ {\rm erf}(x) $ is the error function. We can easily know that the $ \gamma _{auto}(r) $ is dimensionless from Eq.(\ref{gammaauto}) and (\ref{g}).

Furthermore, we can obtain the correlation function $ G_0(z) $ (Fig.\ref{fig:fig2}) by evaluating Eq.(\ref{G}),
\begin{eqnarray}\label{G0}
	G_0(z) = \frac{1}{\ln 2} \left[ f(\zeta) - f\left( \frac{\zeta}{\sqrt{2}}\right)  \right] 
\end{eqnarray}	
where, $ \zeta = z/(2R_g) $ and 
\begin{eqnarray}\label{eq_f}
	f(x) = \exp (-x^2) + (x^2+1) ~{\rm Ei}(-x^2).
\end{eqnarray}
Here, $ {\rm Ei}(x) $ is the exponential integral function defined as \cite{abramowitz1972handbook}
\begin{eqnarray}
		{\rm Ei}(x) = \int_{-\infty}^{x} \frac{e^t}{t} ~dt
\end{eqnarray}
for real non-zero values of $ x $.

Last, we can get $ G(0) = 2\ln (2) A R_g$ by integrating over $ \gamma(r) $, so that Eq.(\ref{P}) can be rewritten as:
\begin{eqnarray}
	P(z) = \exp \left\lbrace 2\ln (2) AR_g \frac{\lambda^2}{S} \left[ G_0(z) -1 \right] \right\rbrace .
\end{eqnarray}
Figure \ref{fig:fig2} clearly shows $ P(0)=1 $ and $ P(z) = \exp \left[  -2\ln(2)AR_g\lambda^2/S \right]    $ is just a constant that is independent of $ G(z) $ when $ z $ tends to infinity because $ G_0(\infty) = 0 $. $ P(\infty) $ is corresponding to the base line of SESANS spectra at large $ z $.

\section{CONCLUSIONS}
This paper assesses how the spin-echo small angle neutron scattering technique, combined with the contrast variation method, may be used to explore the structural properties of polymer chains. Via theoretical calculations of Gaussian chains model system, we have shown that mathematical expressions of the Debye correlation function $ \gamma(r) $, the SESANS spatial correlation $ G(z) $ function and the neutron polarizability function $ P(z) $ analytically. Therefore, SESANS can be used directly for the study of polymer structures without considering the Fourier transform of between $ \gamma_{auto}(r) $ and form factor $ F(Q) $. It is convenient to process experimental data if you use these functions. And SESANS thus can be called the real spatial scattering technique because the analysis of the reciprocal space is not necessary.

\appendix

\section{The Correlation Function of The Density Distribution}\label{appendix1}
From the theoretical aspect, in a molten polymer system which contains hydric and deuterated chains, the fluctuation of the SLD can be derived: 
\begin{eqnarray}\label{a_eq_delt_rho1}
\Delta\rho(\bm{r}')~=&& \rho(\bm{r}')-\bar{\rho}\nonumber\\
=&& b_H\sum_{i=1}^{N_H} \delta(\bm{r}'-\bm{r}'_i) + b_D\sum_{j=1}^{N_D} \delta(\bm{r}'-\bm{r}'_j)-\bar{\rho}~,
\end{eqnarray}
where $ N_H $ and $ N_D $ denote the total number of hydric and deuterated monomers in the system respectively. And the average SLD can be explicitly split into two parts,
\begin{eqnarray}\label{a_eq_rho_bar}
	\bar{\rho}~=~\frac{N_Hb_H+N_Db_D}{V}~=~\bar{\rho}_H+\bar{\rho}_D~.
\end{eqnarray}
Eq.(\ref{a_eq_rho_bar}) is substituted into Eq.(\ref{a_eq_delt_rho1}), then $ \Delta\rho(\bm{r}') $ is rewritten as
\begin{eqnarray}
	\Delta\rho(\bm{r}')~=&& \left[b_H\sum_{i=1}^{N_H} \delta(\bm{r}'-\bm{r}'_i)-\bar{\rho}_H \right] +\left[b_D\sum_{j=1}^{N_D} \delta(\bm{r}'-\bm{r}'_j)-\bar{\rho}_D \right]\nonumber\\
	=&& b_H \left[n_H(\bm{r}')-\bar{n}_H] + b_D[n_D(\bm{r}')-\bar{n}_D\right]\nonumber\\
	=&& b_H \Delta n_H(\bm{r}') + b_D\Delta n_D(\bm{r}')\label{a_eq_delt_rho2}
\end{eqnarray}
where $ n(\bm{r}') $ is the local number density and $ \bar{n} $ is the average number density. Similarly, we can obtain
\begin{eqnarray}\label{a_eq_delt_rho3}
	\Delta\rho(\bm{r}' + \bm{r})~=~b_H \Delta n_H(\bm{r}' + \bm{r}) + b_D\Delta n_D(\bm{r}' + \bm{r})~.
\end{eqnarray}
At the spatial scale of small-angle scattering, the particle number density of the monomer satisfies 
\begin{eqnarray}\label{a_eq_hd0}
	\Delta n_H(\bm{r}')+\Delta n_D(\bm{r}')=0~,
\end{eqnarray}
From Eq.(\ref{a_eq_hd0}), Eq.(\ref{a_eq_delt_rho2}), Eq.(\ref{a_eq_delt_rho3}) and Eq.(\ref{3_eq_difine_gamma}), $ \gamma(\bm{r}) $ is calculated as
\begin{eqnarray}
	\gamma(\bm{r})=&&\left( b_H-b_D\right)^2 \int \left[ \left\langle n_H(\bm{r}') n_H(\bm{r}' + \bm{r})\right\rangle -\bar{n}^2_H\right]  d\bm{r}'\label{a_eq_n_nbarH} \\
	=&&\left( b_H-b_D\right)^2 \int \left[ \left\langle n_D(\bm{r}') n_D(\bm{r}' + \bm{r})\right\rangle -\bar{n}^2_D\right]  d\bm{r}' ~. \label{a_eq_n_nbarD} 
\end{eqnarray}
In Eq.(\ref{a_eq_n_nbarH}), $ \left\langle n_H(\bm{r}') n_H(\bm{r}' + \bm{r}) \right\rangle $ can be expressed in detail as
\begin{eqnarray}
	&&\left\langle n_H(\bm{r}') n_H(\bm{r}' + \bm{r}) \right\rangle \nonumber\\
	=&&\sum_{i=1}^{N_H}\sum_{j=1}^{N_H} \left\langle  \delta(\bm{r}'-\bm{r}'_i)\delta(\bm{r}' + \bm{r}-\bm{r}'_j)  \right\rangle  \nonumber\\
	=&&\sum_{p=1}^{N_{Hc}}\sum_{q=1}^{N_{Hc}}\sum_{m=1}^{N}\sum_{n=1}^{N} \left\langle  \delta(\bm{r}'-\bm{x}_p-\bm{r}_n)\delta(\bm{r}' + \bm{r}-\bm{x}_q-\bm{r}_m)  \right\rangle ~,\label{a_eq_nn1}
\end{eqnarray}
where $ \bm{x} $ denotes the position of the centre of mass a chain, and $ \bm{r}_n $, $ \bm{r}_m $ are displacements of a monomer relative to the barycenter. Here, we split Eq.(\ref{a_eq_nn1}) into $ p=q $ and $ p \neq q $ two parts, 
\begin{eqnarray}
	\left\langle n_H(\bm{r}') n_H(\bm{r}' + \bm{r}) \right\rangle =&&N_{Hc}\sum_{m=1}^{N}\sum_{n=1}^{N} \left\langle  \delta(\bm{r}'-\bm{r}_n)\delta(\bm{r}' + \bm{r}-\bm{r}_m)  \right\rangle \nonumber\\
	 &&+N_{Hc}^2\sum_{m=1}^{N}\sum_{n=1}^{N} \left\langle  \delta(\bm{r}'-\bm{x}_p-\bm{r}_n)\delta(\bm{r}' + \bm{r}-\bm{x}_q-\bm{r}_m)  \right\rangle ~.
\end{eqnarray}
We let 
\begin{eqnarray}\label{a_eq_y1}
	y_1 = \sum_{m=1}^{N}\sum_{n=1}^{N} \left\langle  \delta(\bm{r}'-\bm{r}_n)\delta(\bm{r}' + \bm{r}-\bm{r}_m)  \right\rangle ~,
\end{eqnarray}
and 
\begin{eqnarray}
	y_2 = \sum_{m=1}^{N}\sum_{n=1}^{N} \left\langle  \delta(\bm{r}'-\bm{x}_p-\bm{r}_n)\delta(\bm{r}' + \bm{r}-\bm{x}_q-\bm{r}_m)  \right\rangle ~.
\end{eqnarray}
That means, 
\begin{eqnarray}\label{a_eq_nnH}
	\left\langle n_H(\bm{r}') n_H(\bm{r}' + \bm{r}) \right\rangle -\bar{n}_H^2 = N_{Hc}~y_1 + N_{Hc}^2~ y_2-\bar{n}_H^2 ~,
\end{eqnarray}
\begin{eqnarray}\label{a_eq_nnD}
	\left\langle n_D(\bm{r}') n_D(\bm{r}' + \bm{r}) \right\rangle -\bar{n}_D^2 = N_{Dc}~y_1 + N_{Dc}^2~ y_2-\bar{n}_D^2 ~.
\end{eqnarray}
Using the equivalence of Eq.(\ref{a_eq_nnH}) and (\ref{a_eq_nnD}), we can prove that,
\begin{eqnarray}
	y_2=-\frac{y_1}{N_c}+\frac{N^2}{V^2} ~.
\end{eqnarray}
So the $ \gamma(\bm{r}) $ is simplified as
\begin{eqnarray}\label{a_eq_gamma_y1}
	\gamma(\bm{r}) = N_c \phi (1-\phi) (b_H-b_D)^2  \int y_1 d\bm{r}' ~.
\end{eqnarray}
Besides, note that $ y_1 $ depends only on the form of a chain, so using the first monomer position as the origin of coordinates will simplify the above integral very well. From Eq.(\ref{twounitsdistribution}), we know the possibility density distribution of the $ n $th monomer relative to the first one is 
\begin{eqnarray}\label{a_eq_mono_Gaussian}
	\Phi(\bm{r}_n)~=~\left(\frac{1}{2\pi \sigma_n^2 } \right)^{\frac{3}{2}} \exp \left(  -\frac{|\bm{r}_n|^2}{2\sigma_n^2 }\right),~~~~~	\sigma_n^2 =\frac{nb^2}{3} ~.
\end{eqnarray}
In Eq.(\ref{a_eq_y1}), $ \left\langle ~ \right\rangle  $ denotes the average of all possible conformations of a chain. It can be calculated with Eq.(\ref{a_eq_mono_Gaussian}), 
\begin{eqnarray}
	y_1 = \sum_{n=1}^{N}\sum_{m=1}^{N} \int \delta(\bm{r}'-\bm{r}_n)\delta(\bm{r}' + \bm{r}-\bm{r}_m) \Phi(\bm{r}_n) \Phi(\bm{r}_m) d\bm{r}_n d\bm{r}_m  ~~.
\end{eqnarray}
Therefore, using the previous equation, Eq.(\ref{a_eq_gamma_y1}) can be analytically caculated as
\begin{eqnarray}\label{a_eq_gamma_last}
	\gamma (r) = N_c \phi (1-\phi) (b_H-b_D)^2 \frac{N^2}{4\pi R_g^3} \left[ g(\eta) - \frac{1}{\sqrt{2}}~g\left( \frac{\eta}{\sqrt{2}} \right) - \eta \right] ~,
\end{eqnarray}
where $R_g^2=Nb^2/6~$ \cite{doi1988theory} and $ \eta = r/(2R_g)$. $ \gamma (r) $ only depends on the form of a single chain, so we let $ A = N_c \phi (1-\phi)(b_H - b_D)^2 N^2/(4\pi R^3_g)~ $ and define a dimensionless equation
\begin{eqnarray}
	\gamma _{auto}(r) = g(\eta) - \frac{1}{\sqrt{2}}~g\left( \frac{\eta}{\sqrt{2}} \right) - \eta ~.
\end{eqnarray}

\section{Spatial Correlation Function G(z)}\label{appendix2}
$ G(z) $ is defined by Eq.(\ref{G}). We use Eq.(\ref{a_eq_gamma_last}) and Eq.(\ref{G}) to obtain 
\begin{eqnarray}
		G(z) = 2A \int_{z}^{\infty}   \frac{ \left[ g(\eta) - \frac{1}{\sqrt{2}}~g\left( \frac{\eta}{\sqrt{2}} \right) - \eta \right] r }{\sqrt{r^2-z^2}}  dr ~.
\end{eqnarray}
This integral function will be simplified if we let $ \zeta = z/(2R_g) $,
\begin{eqnarray}
	G(z) =&& 4AR_g \int_{\zeta}^{\infty}   \frac{ \left[ g(\eta) - \frac{1}{\sqrt{2}}~g\left( \frac{\eta}{\sqrt{2}} \right) - \eta \right] \eta }{\sqrt{\eta^2-\zeta^2}}  d\eta \nonumber \\
	=&& 2AR_g \left[ f\left( \zeta \right) - f\left( \frac{\zeta}{\sqrt{2}} \right)  \right] ~.
\end{eqnarray}
Function $ f(x) $ is the Eq.(\ref{eq_f}). And the correlation length will be deduced by
\begin{eqnarray}
	G(0) = \lim_{z \rightarrow 0} G(z) = 2\ln(2)AR_g ~.
\end{eqnarray}
So the normalized SESANS spatial correlation function $ G_0(z) $ can be expressed as
\begin{eqnarray}
	G_0(z) = \frac{G(z)}{G(0)} = \frac{1}{\ln 2}\left[f\left(\frac{z}{2R_g} \right) -f\left(\frac{z}{2\sqrt{2}R_g} \right) \right] ~.
\end{eqnarray}

\bibliography{retouch_manuscript_sesans_polymer}

%aipnum4-2.bst 2019-01-14 (MD) hand-edited version of apsrev4-1.bst
\providecommand{\noopsort}[1]{}\providecommand{\singleletter}[1]{#1}%
\begin{thebibliography}{10}%
\makeatletter
\providecommand \@ifxundefined [1]{%
 \@ifx{#1\undefined}
}%
\providecommand \@ifnum [1]{%
 \ifnum #1\expandafter \@firstoftwo
 \else \expandafter \@secondoftwo
 \fi
}%
\providecommand \@ifx [1]{%
 \ifx #1\expandafter \@firstoftwo
 \else \expandafter \@secondoftwo
 \fi
}%
\providecommand \natexlab [1]{#1}%
\providecommand \enquote  [1]{``#1''}%
\providecommand \bibnamefont  [1]{#1}%
\providecommand \bibfnamefont [1]{#1}%
\providecommand \citenamefont [1]{#1}%
\providecommand \href@noop [0]{\@secondoftwo}%
\providecommand \href [0]{\begingroup \@sanitize@url \@href}%
\providecommand \@href[1]{\@@startlink{#1}\@@href}%
\providecommand \@@href[1]{\endgroup#1\@@endlink}%
\providecommand \@sanitize@url [0]{\catcode `\\12\catcode `\$12\catcode
  `\&12\catcode `\#12\catcode `\^12\catcode `\_12\catcode `\%12\relax}%
\providecommand \@@startlink[1]{}%
\providecommand \@@endlink[0]{}%
\providecommand \url  [0]{\begingroup\@sanitize@url \@url }%
\providecommand \@url [1]{\endgroup\@href {#1}{\urlprefix }}%
\providecommand \urlprefix  [0]{URL }%
\providecommand \Eprint [0]{\href }%
\providecommand \doibase [0]{https://doi.org/}%
\providecommand \selectlanguage [0]{\@gobble}%
\providecommand \bibinfo  [0]{\@secondoftwo}%
\providecommand \bibfield  [0]{\@secondoftwo}%
\providecommand \translation [1]{[#1]}%
\providecommand \BibitemOpen [0]{}%
\providecommand \bibitemStop [0]{}%
\providecommand \bibitemNoStop [0]{.\EOS\space}%
\providecommand \EOS [0]{\spacefactor3000\relax}%
\providecommand \BibitemShut  [1]{\csname bibitem#1\endcsname}%
\let\auto@bib@innerbib\@empty
%</preamble>
\bibitem [{\citenamefont {Kawakatsu}(2004)}]{kawakatsu2004statistical}%
  \BibitemOpen
  \bibfield  {author} {\bibinfo {author} {\bibfnamefont {T.}~\bibnamefont
  {Kawakatsu}},\ }\href@noop {} {\emph {\bibinfo {title} {Statistical physics
  of polymers: an introduction}}}\ (\bibinfo  {publisher} {Springer Science \&
  Business Media},\ \bibinfo {year} {2004})\BibitemShut {NoStop}%
\bibitem [{\citenamefont {Bouwman}\ \emph {et~al.}(2000)\citenamefont
  {Bouwman}, \citenamefont {Oossanen}, \citenamefont {Uca}, \citenamefont
  {Kraan},\ and\ \citenamefont {Rekveldt}}]{bouwman2000development}%
  \BibitemOpen
  \bibfield  {author} {\bibinfo {author} {\bibfnamefont {W.~G.}\ \bibnamefont
  {Bouwman}}, \bibinfo {author} {\bibfnamefont {M.~v.}\ \bibnamefont
  {Oossanen}}, \bibinfo {author} {\bibfnamefont {O.}~\bibnamefont {Uca}},
  \bibinfo {author} {\bibfnamefont {W.~H.}\ \bibnamefont {Kraan}},\ and\
  \bibinfo {author} {\bibfnamefont {M.~T.}\ \bibnamefont {Rekveldt}},\
  }\bibfield  {title} {\enquote {\bibinfo {title} {Development of spin-echo
  small-angle neutron scattering},}\ }\href@noop {} {\bibfield  {journal}
  {\bibinfo  {journal} {Journal of applied crystallography}\ }\textbf {\bibinfo
  {volume} {33}},\ \bibinfo {pages} {767--770} (\bibinfo {year}
  {2000})}\BibitemShut {NoStop}%
\bibitem [{\citenamefont {Doi}\ and\ \citenamefont
  {Edwards}(1988)}]{doi1988theory}%
  \BibitemOpen
  \bibfield  {author} {\bibinfo {author} {\bibfnamefont {M.}~\bibnamefont
  {Doi}}\ and\ \bibinfo {author} {\bibfnamefont {S.~F.}\ \bibnamefont
  {Edwards}},\ }\href@noop {} {\emph {\bibinfo {title} {The theory of polymer
  dynamics}}},\ Vol.~\bibinfo {volume} {73}\ (\bibinfo  {publisher} {oxford
  university press},\ \bibinfo {year} {1988})\BibitemShut {NoStop}%
\bibitem [{\citenamefont {De~Gennes}(1979)}]{de1979scaling}%
  \BibitemOpen
  \bibfield  {author} {\bibinfo {author} {\bibfnamefont {P.-G.}\ \bibnamefont
  {De~Gennes}},\ }\href@noop {} {\emph {\bibinfo {title} {Scaling concepts in
  polymer physics}}}\ (\bibinfo  {publisher} {Cornell university press},\
  \bibinfo {year} {1979})\BibitemShut {NoStop}%
\bibitem [{\citenamefont {Krouglov}\ \emph
  {et~al.}(2003{\natexlab{a}})\citenamefont {Krouglov}, \citenamefont
  {De~Schepper}, \citenamefont {Bouwman},\ and\ \citenamefont
  {Rekveldt}}]{krouglov2003real}%
  \BibitemOpen
  \bibfield  {author} {\bibinfo {author} {\bibfnamefont {T.}~\bibnamefont
  {Krouglov}}, \bibinfo {author} {\bibfnamefont {I.~M.}\ \bibnamefont
  {De~Schepper}}, \bibinfo {author} {\bibfnamefont {W.~G.}\ \bibnamefont
  {Bouwman}},\ and\ \bibinfo {author} {\bibfnamefont {M.~T.}\ \bibnamefont
  {Rekveldt}},\ }\bibfield  {title} {\enquote {\bibinfo {title} {Real-space
  interpretation of spin-echo small-angle neutron scattering},}\ }\href@noop {}
  {\bibfield  {journal} {\bibinfo  {journal} {Journal of applied
  crystallography}\ }\textbf {\bibinfo {volume} {36}},\ \bibinfo {pages}
  {117--124} (\bibinfo {year} {2003}{\natexlab{a}})}\BibitemShut {NoStop}%
\bibitem [{\citenamefont {Kruglov}(2005)}]{kruglov2005correlation}%
  \BibitemOpen
  \bibfield  {author} {\bibinfo {author} {\bibfnamefont {T.}~\bibnamefont
  {Kruglov}},\ }\bibfield  {title} {\enquote {\bibinfo {title} {Correlation
  function of the excluded volume},}\ }\href@noop {} {\bibfield  {journal}
  {\bibinfo  {journal} {Journal of applied crystallography}\ }\textbf {\bibinfo
  {volume} {38}},\ \bibinfo {pages} {716--720} (\bibinfo {year}
  {2005})}\BibitemShut {NoStop}%
\bibitem [{\citenamefont {Andersson}\ \emph {et~al.}(2008)\citenamefont
  {Andersson}, \citenamefont {Van~Heijkamp}, \citenamefont {De~Schepper},\ and\
  \citenamefont {Bouwman}}]{andersson2008analysis}%
  \BibitemOpen
  \bibfield  {author} {\bibinfo {author} {\bibfnamefont {R.}~\bibnamefont
  {Andersson}}, \bibinfo {author} {\bibfnamefont {L.~F.}\ \bibnamefont
  {Van~Heijkamp}}, \bibinfo {author} {\bibfnamefont {I.~M.}\ \bibnamefont
  {De~Schepper}},\ and\ \bibinfo {author} {\bibfnamefont {W.~G.}\ \bibnamefont
  {Bouwman}},\ }\bibfield  {title} {\enquote {\bibinfo {title} {Analysis of
  spin-echo small-angle neutron scattering measurements},}\ }\href@noop {}
  {\bibfield  {journal} {\bibinfo  {journal} {Journal of Applied
  Crystallography}\ }\textbf {\bibinfo {volume} {41}},\ \bibinfo {pages}
  {868--885} (\bibinfo {year} {2008})}\BibitemShut {NoStop}%
\bibitem [{\citenamefont {Li}\ \emph {et~al.}(2010)\citenamefont {Li},
  \citenamefont {Shew}, \citenamefont {Liu}, \citenamefont {Pynn},
  \citenamefont {Liu}, \citenamefont {Herwig}, \citenamefont {Smith},
  \citenamefont {Robertson},\ and\ \citenamefont {Chen}}]{li2010theoretical}%
  \BibitemOpen
  \bibfield  {author} {\bibinfo {author} {\bibfnamefont {X.}~\bibnamefont
  {Li}}, \bibinfo {author} {\bibfnamefont {C.-Y.}\ \bibnamefont {Shew}},
  \bibinfo {author} {\bibfnamefont {Y.}~\bibnamefont {Liu}}, \bibinfo {author}
  {\bibfnamefont {R.}~\bibnamefont {Pynn}}, \bibinfo {author} {\bibfnamefont
  {E.}~\bibnamefont {Liu}}, \bibinfo {author} {\bibfnamefont {K.~W.}\
  \bibnamefont {Herwig}}, \bibinfo {author} {\bibfnamefont {G.~S.}\
  \bibnamefont {Smith}}, \bibinfo {author} {\bibfnamefont {J.~L.}\ \bibnamefont
  {Robertson}},\ and\ \bibinfo {author} {\bibfnamefont {W.-R.}\ \bibnamefont
  {Chen}},\ }\bibfield  {title} {\enquote {\bibinfo {title} {Theoretical
  studies on the structure of interacting colloidal suspensions by spin-echo
  small angle neutron scattering},}\ }\href@noop {} {\bibfield  {journal}
  {\bibinfo  {journal} {The Journal of chemical physics}\ }\textbf {\bibinfo
  {volume} {132}},\ \bibinfo {pages} {174509} (\bibinfo {year}
  {2010})}\BibitemShut {NoStop}%
\bibitem [{\citenamefont {Krouglov}\ \emph
  {et~al.}(2003{\natexlab{b}})\citenamefont {Krouglov}, \citenamefont
  {Bouwman}, \citenamefont {Plomp}, \citenamefont {Rekveldt}, \citenamefont
  {Vroege}, \citenamefont {Petukhov},\ and\ \citenamefont
  {Thies-Weesie}}]{krouglov2003structural}%
  \BibitemOpen
  \bibfield  {author} {\bibinfo {author} {\bibfnamefont {T.}~\bibnamefont
  {Krouglov}}, \bibinfo {author} {\bibfnamefont {W.~G.}\ \bibnamefont
  {Bouwman}}, \bibinfo {author} {\bibfnamefont {J.}~\bibnamefont {Plomp}},
  \bibinfo {author} {\bibfnamefont {M.~T.}\ \bibnamefont {Rekveldt}}, \bibinfo
  {author} {\bibfnamefont {G.~J.}\ \bibnamefont {Vroege}}, \bibinfo {author}
  {\bibfnamefont {A.~V.}\ \bibnamefont {Petukhov}},\ and\ \bibinfo {author}
  {\bibfnamefont {D.~M.}\ \bibnamefont {Thies-Weesie}},\ }\bibfield  {title}
  {\enquote {\bibinfo {title} {Structural transitions of hard-sphere colloids
  studied by spin-echo small-angle neutron scattering},}\ }\href@noop {}
  {\bibfield  {journal} {\bibinfo  {journal} {Journal of Applied
  Crystallography}\ }\textbf {\bibinfo {volume} {36}},\ \bibinfo {pages}
  {1417--1423} (\bibinfo {year} {2003}{\natexlab{b}})}\BibitemShut {NoStop}%
\bibitem [{\citenamefont {Abramowitz}\ and\ \citenamefont
  {Stegun}(1972)}]{abramowitz1972handbook}%
  \BibitemOpen
  \bibfield  {author} {\bibinfo {author} {\bibfnamefont {M.}~\bibnamefont
  {Abramowitz}}\ and\ \bibinfo {author} {\bibfnamefont {I.~A.}\ \bibnamefont
  {Stegun}},\ }\bibfield  {title} {\enquote {\bibinfo {title} {Handbook of
  mathematical functions with formulas, graphs, and mathematical tables},}\
  }\href@noop {} {\  (\bibinfo {year} {1972})}\BibitemShut {NoStop}%
\end{thebibliography}%

\end{document}